\documentclass{cimento}

\usepackage{graphicx}

\title{GRB 050315: A step toward the uniqueness of the overall GRB structure and the true nature of long GRBs}
\author{R.~Ruffini\from{icra}\from{dipfis}\thanks{ruffini@icra.it}\ETC,
M.G.~Bernardini\from{icra}\from{dipfis},
C.L.~Bianco\from{icra}\from{dipfis},
P.~Chardonnet\from{icra}\from{savoie},
F.~Fraschetti\from{icra}\from{brera},
R.~Guida\from{icra}\from{dipfis},
        \atque
S.-S.~Xue\from{icra}\from{dipfis}}

\instlist{
  \inst{icra} ICRANet and ICRA, Piazzale della Repubblica 10, I--65100 Pescara, Italy.
  \inst{dipfis} Dipartimento di Fisica, Universit\`a di Roma ``La Sapienza'', Piazzale Aldo Moro 5, I-00185 Roma, Italy.
  \inst{savoie} Universit\'e de Savoie, LAPTH - LAPP, BP 110, F-74941 Annecy-le-Vieux Cedex, France.
  \inst{brera} Osservatorio Astronomico di Brera, via Bianchi 46, I-23807 Merate (LC), Italy.
  }
\PACSes{\PACSit{98.70.Rz}{gamma-ray sources; gamma-ray bursts}}

\begin{document}

\maketitle

\begin{abstract}
Using the \emph{Swift} data of GRB 050315, we progress on the uniqueness of our theoretically predicted Gamma-Ray Burst (GRB) structure as composed by a proper-GRB (P-GRB), emitted at the transparency of an electron-positron plasma with suitable baryon loading, and an afterglow comprising the so called ``prompt emission'' as due to external shocks. Thanks to the \emph{Swift} observations, we can theoretically fit detailed light curves for selected energy bands on a continuous time scale ranging over $10^6$ seconds. The theoretically predicted instantaneous spectral distribution over the entire afterglow confirms a clear hard-to-soft behavior encompassing, continuously, the ``prompt emission'' all the way to the latest phases of the afterglow. Consequences of the instrumental threshold on the definition of ``short'' and ``long'' GRBs are discussed.
\end{abstract}

\section{Introduction}

GRB 050315 \cite{va05} has been triggered and located by the BAT instrument \cite{b04,ba05} on board of the {\em Swift} satellite \cite{ga04} at 2005-March-15 20:59:42 UT \cite{pa05}. The narrow field instrument XRT \cite{bua04,bua05} began observations $\sim 80$ s after the BAT trigger, one of the earliest XRT observations yet made, and continued to detect the source for $\sim 10$ days \cite{va05}. The spectroscopic redshift has been found to be $z = 1.949$ \cite{kb05}.

We summarize the results, recently published \cite{050315}, of the fit of the \emph{Swift} data of this source in $5$ energy bands in the framework of our theoretical model (see \cite{rlet1,rlet2,r02,rubr,rubr2,EQTS_ApJL,EQTS_ApJL2,PowerLaws} and references therein). We here point out a new step toward the uniqueness of the explanation of the overall GRB structure and consequences on the definition of the ``short'' and ``long'' GRBs. In this respect, we emphasize the essential role of the instrumental threshold.

\section{Our theoretical model}\label{model}

A basic feature of our model consists in a sharp distinction between two different components in the GRB structure: {\bf 1)} the Proper-GRB (P-GRB), emitted at the moment of transparency of the self-accelerating $e^\pm$-baryons plasma (see e.g. \cite{g86,p86,sp90,psn93,mlr93,gw98,rswx99,rswx00,rlet1,rlet2,Monaco_RateEq}); {\bf 2)} an afterglow described by the interaction with the interstellar medium (ISM) of the baryons accelerated during the optically thick phase before the P-GRB emission (see \cite{rswx99,rswx00,rlet2,rubr} and references therein). Such an afterglow is composed of three different regimes. The first afterglow regime corresponds to a bolometric luminosity monotonically increasing with the photon detector arrival time, corresponding to a substantially constant Lorentz gamma factor of the accelerated baryons. The second regime consists of the bolometric luminosity peak, corresponding to the ``knee'' in the decreasing phase of the baryonic Lorentz gamma factor. The third regime corresponds to a bolometric luminosity decreasing with arrival time, corresponding to the late deceleration of the Lorentz gamma factor (see \cite{rlet2,rubr} for details). What is usually called ``prompt emission'' in our case coincides with the peak of the afterglow emission. In \cite{rlet2} we have chosen as a prototype the source GRB 991216 which clearly shows the existence of the P-GRB and the three regimes of the afterglow. Both the relative intensity of the P-GRB to the peak of the afterglow and their corresponding temporal lag were theoretically predicted within a few percent (see Fig. 11 in \cite{rubr}). Unfortunately, data from BATSE existed only up to $36$ s, and data from R-XTE and Chandra only after $3500$ s, leaving our theoretical predictions in the whole range between $36$ s and $3500$ s without the support of the comparison with observational data.

\begin{figure}
\includegraphics[width=0.5\hsize,clip]{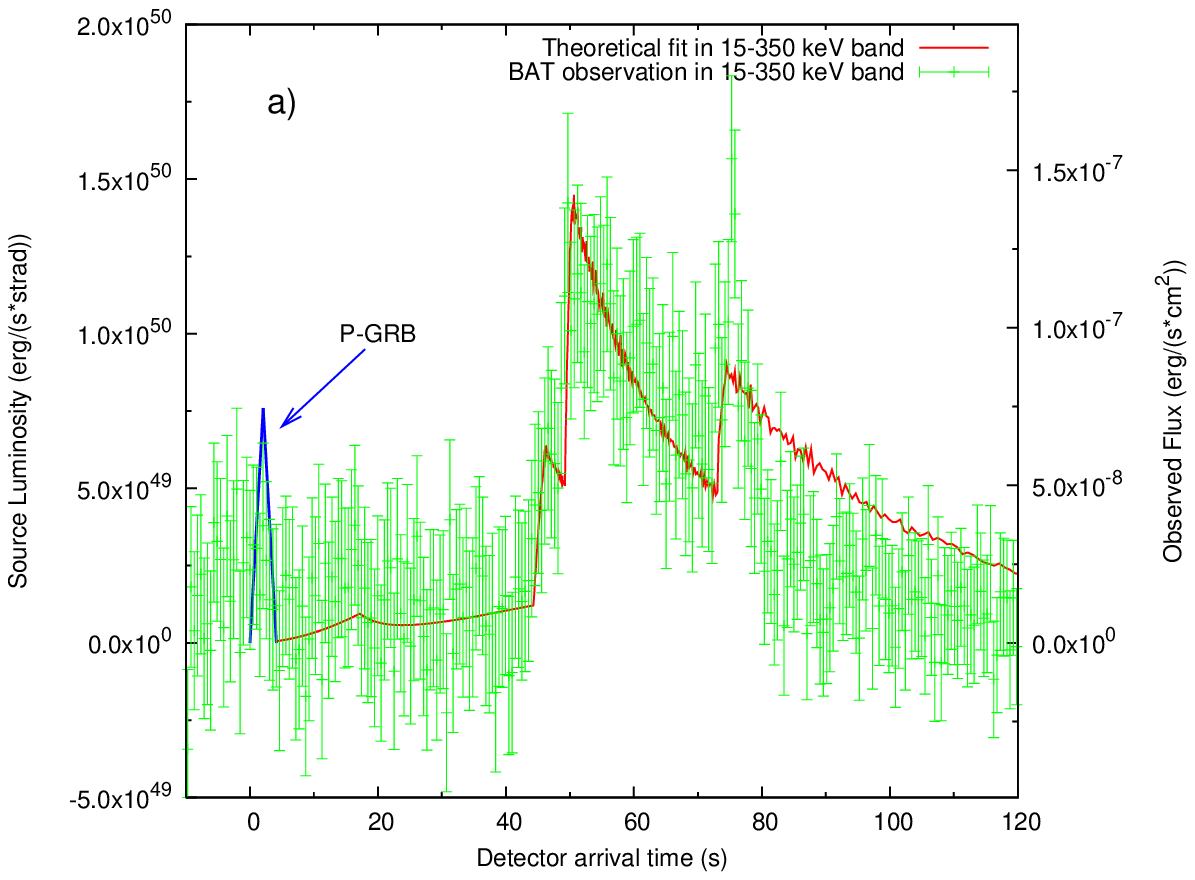}
\includegraphics[width=0.5\hsize,clip]{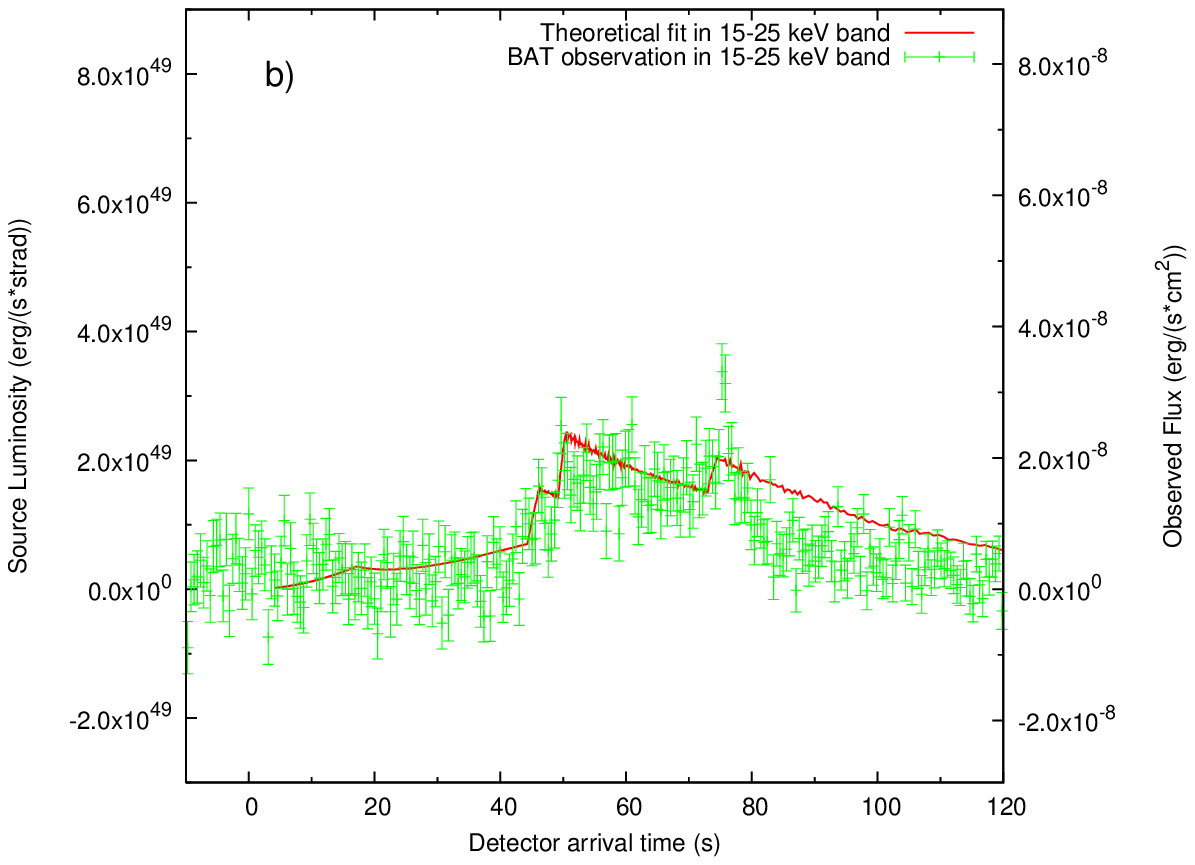}\\
\includegraphics[width=0.5\hsize,clip]{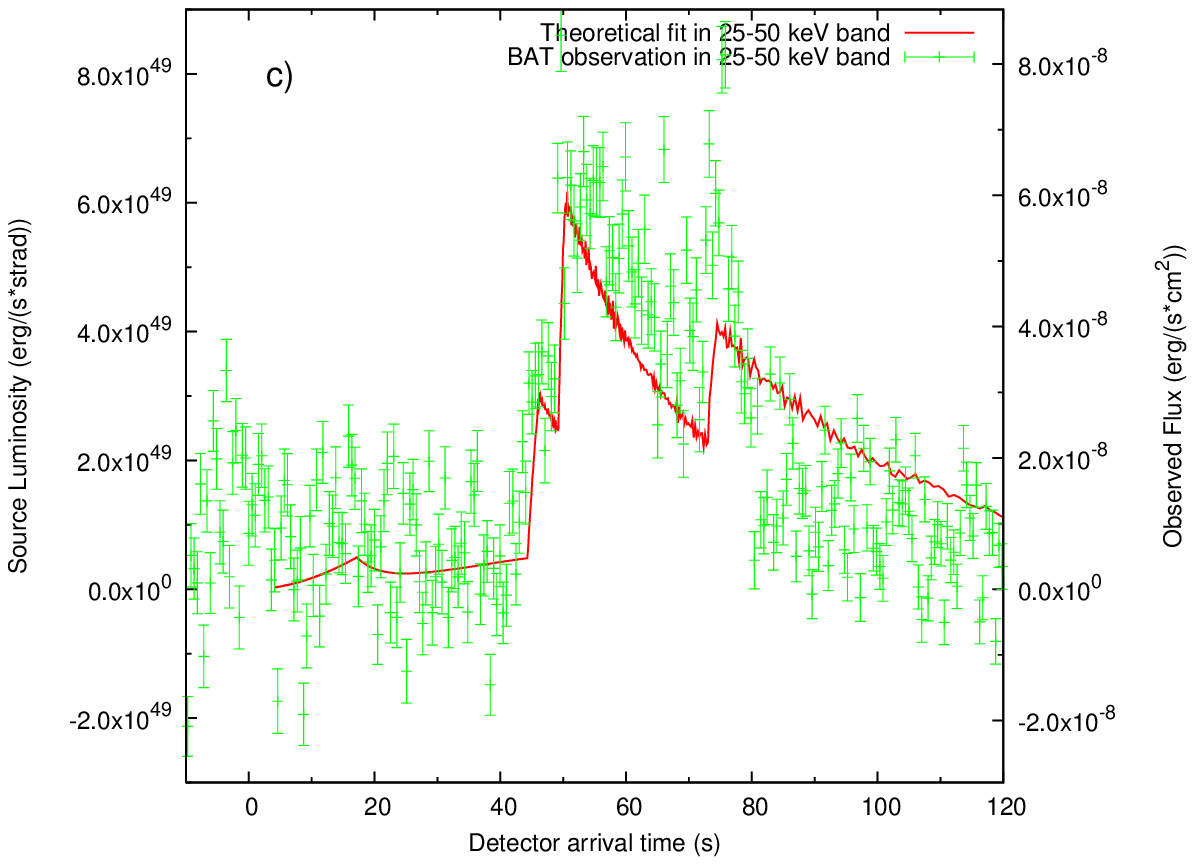}
\includegraphics[width=0.5\hsize,clip]{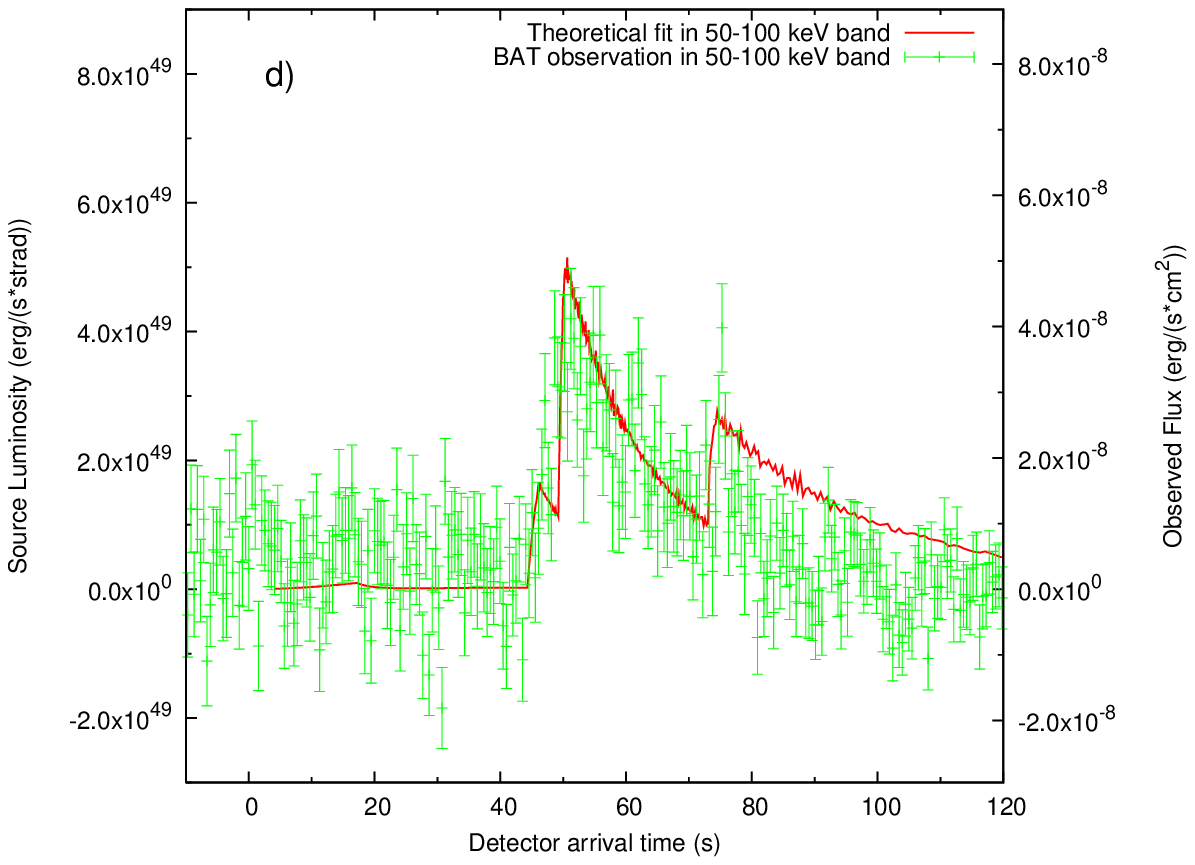}
\caption{Our theoretical fit (red line) of the BAT observations (green points) of GRB 050315 in the $15$--$350$ keV (a), $15$--$25$ keV (b), $25$--$50$ keV (c), $50$--$100$ keV (d) energy bands \cite{va05}. The blue line in panel (a) represents our theoretical prediction for the intensity and temporal position of the P-GRB.}
\label{tot}
\end{figure}

The consistency of our model has been tested in a variety of other sources, like GRB 980425 \cite{cospar02}, GRB 030329 \cite{mg10grazia}, GRB 031203 \cite{031203}. Thanks to the data provided by the \emph{Swift} satellite, we are finally able to confirm our theoretical predictions on the GRB structure with a detailed fit of the complete afterglow light curve of GRB 050315, from the peak, including the ``prompt emission'', all the way to the latest phases without any gap in the observational data \cite{050315}. The ``prompt emission'' in our model is not due to the prolonged activity of an ``inner engine'' \cite{rlet2}.

\section{The fit of the observations}\label{fit}

The best fit of the observational data is obtained by the determination of the two free parameters describing the source, namely {\bf 1)} the total energy $E_{e\pm}^{tot}$ of the $e^\pm$ plasma and {\bf 2)} its baryon loading $B \equiv M_Bc^2/E_{e\pm}^{tot}$, as well as of the ISM distribution surrounding the source. Such ISM distribution is characterized by two additional parameters which are function of the distance $r$ from the source: {\bf 1)} the ISM number density $n_{ISM} \left(r\right)$ and {\bf 2)} the $\mathcal{R}\left(r\right)$ parameter which is the ratio between the effective emitting area of the expanding bayonic shell and its total visible area.

We obtain for the first parameter $E_{e\pm}^{tot} = 1.46\times 10^{53}$ erg (the observational \emph{Swift} $E_{iso}$ is $> 2.62\times 10^{52}$ erg, see \cite{va05}), so that the plasma is created between the radii $r_1 = 5.88\times 10^6$ cm and $r_2 = 1.74 \times 10^8$ cm with an initial temperature $T = 2.05 MeV$ and a total number of pairs $N_{e^+e^-} = 7.93\times 10^{57}$. For the second parameter we find $B = 4.55 \times 10^{-3}$. The transparency point and the P-GRB emission occurs then with an initial Lorentz gamma factor of the accelerated baryons $\gamma_\circ = 217.81$ at a distance $r = 1.32 \times 10^{14}$ cm from the black hole.

The ISM average density between the transparency point (i.e. the P-GRB emission) and the beginning of the peak of the afterglow is fixed by the temporal delay between such afterglow peak and the P-GRB. The corresponding ${\cal R}$ parameter value in this initial region, instead, is simply extrapolated backward from the peak of the afterglow. The structure of the ``prompt emission'' has been reproduced assuming three overdense spherical ISM regions with width $\Delta$ and density contrast $\Delta n/\langle n\rangle$: we chose for the first region, at $r = 4.15\times 10^{16}$ cm, $\Delta = 1.5\times 10^{15}$ cm and $\Delta n/\langle n\rangle = 5.17$, for the second region, at $r = 4.53\times 10^{16}$ cm, $\Delta = 7.0\times 10^{14}$ cm and $\Delta n/\langle n\rangle = 36.0$ and for the third region, at $r = 5.62\times 10^{16}$ cm, $\Delta = 5.0\times 10^{14}$ cm and $\Delta n/\langle n\rangle = 85.4$. The ISM mean density during this phase is $\left\langle n_{ISM} \right\rangle=0.81$ particles/cm$^3$ and $\left\langle {\cal R} \right\rangle = 1.4 \times 10^{-7}$. With this choice of the density mask we obtain agreement with the observed light curve, as shown in Fig. \ref{tot}.

\begin{figure}
\includegraphics[width=\hsize,clip]{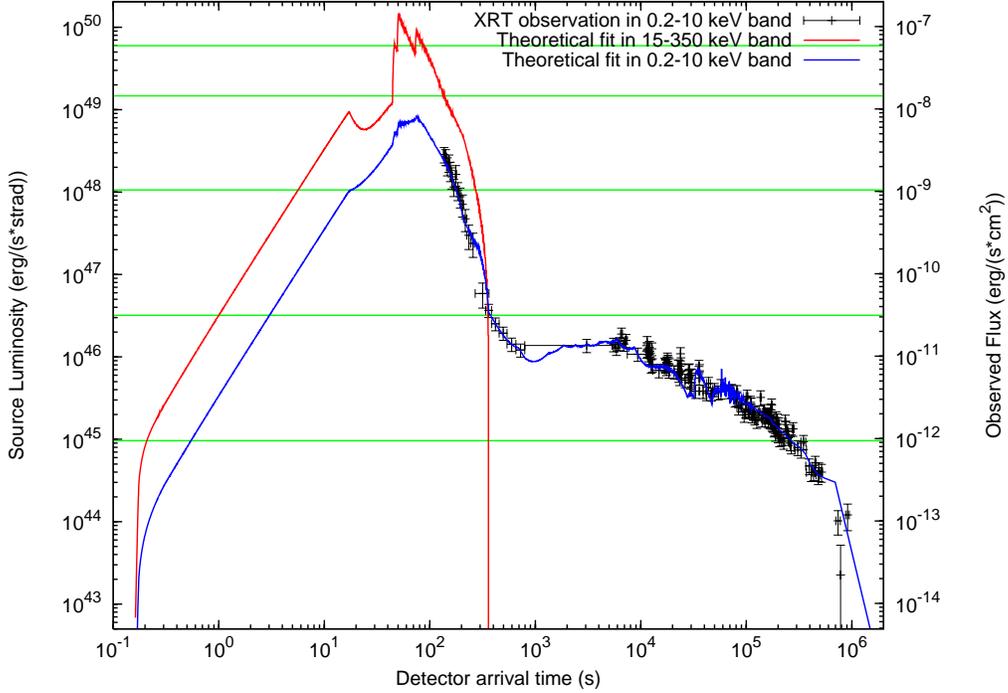}
\caption{Our theoretical fit (blue line) of the XRT observations (black points) of GRB 050315 in the $0.2$--$10$ keV energy band \cite{va05}. The theoretical fit of the BAT observations (see Fig. \ref{tot}a) in the $15$--$350$ keV energy band is also represented (red line). The horizontal green lines corresponds to different possible instrumental thresholds.}
\label{global}
\end{figure}

It has been recently pointed out \cite{meszaros,nousek,panaitescu,zhang} that almost all the GRBs observed by {\em Swift} show a ``canonical behavior'': an initial very steep decay followed by a shallow decay and finally a steeper decay. In our treatment the baryonic shell expands in an ISM region, between $r = 9.00\times 10^{16}$ cm and $r = 5.50\times 10^{18}$ cm, which is significantly at lower density ($\left\langle n_{ISM} \right\rangle=4.76 \times 10^{-4}$ particles/cm$^3$, $\left\langle {\cal R} \right\rangle = 7.0 \times 10^{-6}$) then the one corresponding to the prompt emission, and this produces a slower decrease of the velocity of the baryons with a consequent longer duration of the afterglow emission. The initial steep decay of the observed flux is due to the smaller number of collisions with the ISM. In Fig. \ref{global} is represented our theoretical fit of the XRT data, together with the theoretically computed $15$--$350$ keV light curve of Fig. \ref{tot}a. Both the prompt and the afterglow emission are just due to the thermal radiation in the comoving frame produced by inelastic collisions with the ISM duly boosted by the relativistic transformations over the EQTSs.

\section{Conclusions}\label{concl}

The confirmation, obtained by \emph{Swift}, of our theoretical model, which identifies for every GRB the two distinct component of a P-GRB and an afterglow, implies a revision of the usual classification of the GRB sources. We recall that, following the classification proposed by \cite{ka93} and, later, by \cite{t98}, there are two distinct kind of GRB sources: the so-called ``short GRBs'', lasting less than $\sim 1$ s and with harder spectrum, and the remaining ``long GRBs''. In our \cite{rlet2} we had proposed that the basic mechanism underlying both processes is identical and consists in the $e^\pm$ plasma generated by the vacuum polarization process occurring in the gravitational collapse leading to the formation of a black hole. The difference between these two families of sources is due to the value of the $B$ parameter: for $B < 10^{-5}$ the P-GRB becomes predominant with respect to the afterglow and we observe in such case a ``short GRB''; for larger values of the $B$ parameter, which in any case should be $B < 10^{-2}$, the afterglow is predominant with respect to the P-GRB (see details in \cite{rubr,rubr2}).

The confirmation by \emph{Swift} of our prediction of the overall afterglow structure, and especially the coincidence of the ``prompt emission'' with the peak of the afterglow, opens a new problematic in the definition of the long GRBs. It is clear, in fact, that the identification of the ``prompt emission'' in the current GRB literature is not at all intrinsic to the phenomenon but is merely due to the threshold of the instruments used in the observations (e.g. BATSE in the $50$--$300$ keV energy range, or BeppoSAX GRBM in $40$--$700$ keV, or \emph{Swift} BAT in $15$--$350$ keV). As it is clear from Fig. \ref{global}, there is no natural way to identify in the source a special extension of the peak of the afterglow that is not the one purely defined by the experimental threshold. It is clear, therefore, that long GRBs, as defined till today, are just the peak of the afterglow and there is no way, as explained above, to define their ``prompt emission'' duration as a characteristic signature of the source. As the \emph{Swift} observations show, the duration of the long GRBs has to coincide with the duration of the entire afterglow. A Kouveliotou - Tavani plot of the long GRBs, done following the correct interpretation made possible by the \emph{Swift} data, will present enormous dispersion on the temporal axis.

It is very interesting, however, that the current analysis we are making of the \emph{Swift} observations, especially of GRB 060218, have opened a further possibility for an alternative class of short GRBs, quite different from the ones initially hypothesized in our \cite{rlet2}. It is indeed possible to have sources with very small average ISM density (e.g. $\langle n_{ISM} \rangle < 10^{-3}$) in which, although the baryon loading is much larger than $10^{-4}$, we have in fact the predominance of the peak luminosity of the P-GRB with respect to the one of the afterglow \cite{970228}. This paradox can be indeed simply expressed by the fact that, although the total energy of the afterglow (which depends only on $B$) is much larger than the one of the P-GRB, its luminosity, in view of the very low value of ISM average density, extends on a much longer time scale. Its peak luminosity is therefore much smaller than the one of the P-GRBs.


\begin{thebibliography}{0}

\bibitem{va05}
\BY{Vaughan S. \etal} \IN{ApJ}{638}{2006}{920}.

\bibitem{b04}
\BY{Barthelmy S.D.} \IN{SPIE}{5165}{2004}{175}.

\bibitem{ba05}
\BY{Barthelmy S.D. \etal} \IN{Sp. Sc. Rev.}{120}{2005}{143}.

\bibitem{ga04}
\BY{Gehrels N. \etal} \IN{ApJ}{611}{2004}{1005}.

\bibitem{pa05}
\BY{Parsons A. \etal} \IN{GCN}{3094}{2005}{}.

\bibitem{bua04}
\BY{Burrows D.N. \etal} \IN{SPIE}{5165}{2004}{201}.

\bibitem{bua05}
\BY{Burrows D.N. \etal} \IN{Sp. Sc. Rev.}{120}{2005}{165}.

\bibitem{kb05}
\BY{Kelson D. \atque Berger E.} \IN{GCN}{3101}{2005}{}.

\bibitem{050315}
\BY{Ruffini R., Bernardini M.G., Bianco C.L., Chardonnet P., Fraschetti F., Guida R. \atque Xue S.-S.} \IN{ApJL}{645}{2006}{L109}.

\bibitem{rlet1}
\BY{Ruffini R., Bianco C.L., Chardonnet P., Fraschetti F. \atque Xue S.-S.} \IN{ApJL}{555}{2001a}{L107}.

\bibitem{rlet2}
\BY{Ruffini R., Bianco C.L., Chardonnet P., Fraschetti F. \atque Xue S.-S.} \IN{ApJL}{555}{2001a}{L113}.

\bibitem{r02}
\BY{Ruffini R., Bianco C.L., Chardonnet P., Fraschetti F. \atque Xue S.-S.} \IN{ApJL}{581}{2002}{L19}.

\bibitem{rubr}
\BY{Ruffini R., Bianco C.L., Chardonnet P., Fraschetti F., Vitagliano L. \atque Xue S.-S.} in \TITLE{COSMOLOGY AND GRAVITATION: X$^{th}$ Brazilian School of Cosmology and Gravitation; 25$^{th}$ Anniversary (1977-2002)}, edited by \NAME{Novello M. \atque Perez Bergliaffa S.E.} \IN{AIP Conf. Proc.}{668}{2003}{16}.

\bibitem{rubr2}
\BY{Ruffini R., Bernardini M.G., Bianco C.L., Chardonnet P., Fraschetti F., Gurzadyan V., Vitagliano L. \atque Xue S.-S.} in \TITLE{COSMOLOGY AND GRAVITATION: ${\rm XI}^{th}$ Brazilian School of Cosmology and Gravitation}, edited by \NAME{Novello M. \atque Perez Bergliaffa S.E.} \IN{AIP Conf. Proc.}{782}{2005}{42}.

\bibitem{EQTS_ApJL}
\BY{Bianco C.L. \atque Ruffini R.} \IN{ApJL}{605}{2004}{L1}.

\bibitem{EQTS_ApJL2}
\BY{Bianco C.L. \atque Ruffini R.} \IN{ApJL}{620}{2005}{L23}.

\bibitem{PowerLaws}
\BY{Bianco C.L. \atque Ruffini R.} \IN{ApJL}{633}{2005}{L13}.

\bibitem{g86}
\BY{Goodman J.} \IN{ApJ}{308}{1986}{L47}.

\bibitem{p86}
\BY{Paczy\'nski B.} \IN{ApJ}{308}{1986}{L43}.

\bibitem{sp90}
\BY{Shemi A. \atque Piran T.} \IN{ApJ}{365}{1990}{L55}.

\bibitem{psn93}
\BY{Piran T., Shemi A. \atque Narayan R.} \IN{MNRAS}{263}{1993}{861}.

\bibitem{mlr93}
\BY{M\'esz\'aros P., Laguna P. \atque Rees M.J.} \IN{ApJ}{415}{1993}{181}.

\bibitem{gw98}
\BY{Grimsrud O.M. \atque Wasserman I.} \IN{MNRAS}{300}{1998}{1158}.

\bibitem{rswx99} 
\BY{Ruffini R., Salmonson J.D., Wilson J.R. \atque Xue S.-S.} \IN{A\&A}{350}{1999}{334}.

\bibitem{rswx00} 
\BY{Ruffini R., Salmonson J.D., Wilson J.R. \atque Xue S.-S.} \IN{A\&A}{359}{2000}{855}.

\bibitem{Monaco_RateEq}
\BY{Ruffini R., Bianco C.L., Vereshchagin G. \atque Xue S.-S.}, in \TITLE{Proceedings of the  Relativistic Astrophysics and Cosmology - Einstein's Legacy  meeting}, edited by \NAME{Aschenbach B., Burwitz V., Hasinger G. \atque Leibundgut B.} (Springer-Verlag, in press) 2006.

\bibitem{cospar02}
\BY{Ruffini R., Bianco C.L., Chardonnet P., Fraschetti F. \atque Xue S.-S.} \IN{Adv. Sp. Res.}{34}{2004}{2715}.

\bibitem{mg10grazia}
\BY{Bernardini M.G., Bianco C.L., Chardonnet P., Fraschetti F., Ruffini R. \atque Xue S.-S.} in \TITLE{Proceedings of the tenth Marcel Grossmann Meeting}, edited by \NAME{Novello M. \atque Perez-Bergliaffa S.E.} (World Scientific, Singapore) 2006, p. 2459.

\bibitem{031203}
\BY{Bernardini M.G., Bianco C.L., Chardonnet P., Fraschetti F., Ruffini R. \atque Xue S.-S.} \IN{ApJL}{634}{2005}{L29}.

\bibitem{meszaros}
\BY{M\'esz\'aros, P.} in \TITLE{16th Annual October Astrophysics Conference in Maryland}, edited by \NAME{Holt S., Gehrels N. \atque Nousek J.} \IN{AIP Conf. Proc.}{836}{2006}{234}.

\bibitem{nousek}
\BY{Nousek J.A. \etal} \IN{ApJ}{642}{2006}{389}.

\bibitem{panaitescu}
\BY{Panaitescu A., M\'esz\'aros P., Gehrels N., Burrows D. \atque Nousek J.} \IN{MNRAS}{366}{2006}{1357}.

\bibitem{zhang}
\BY{Zhang B. \etal} \IN{ApJ}{642}{2006}{354}.

\bibitem{ka93}
\BY{Kouveliotou C. \etal} \IN{ApJL}{413}{1993}{L101}.

\bibitem{t98}
\BY{Tavani M.} \IN{ApJL}{497}{1998}{L21}.

\bibitem{970228}
\BY{Bernardini M.G., Bianco C.L., Caito L., Chardonnet P., Corsi A., Dainotti M.G., Fraschetti F., Guida R., Ruffini R. \atque Xue S.-S.} \IN{Il Nuovo Cimento C}{On this same volume}{}{}.

\end{thebibliography}
\end{document}